\documentclass[twocolumn,showpacs,preprintnumbers,amsmath,amssymb,superscriptaddress]{revtex4-1}

\usepackage{graphicx}
\usepackage{dcolumn}
\usepackage{bm}
\usepackage{bigints}
\usepackage{xcolor}  

\usepackage{blindtext}
\usepackage{newfloat}
\DeclareFloatingEnvironment[name={S}]{suppfigure}
\usepackage{graphicx}
\usepackage{dcolumn}
\usepackage{bm}
\usepackage{upgreek}
\usepackage{amsmath}
\usepackage{dsfont}
\usepackage{bigints}
\usepackage{graphicx}
\usepackage{color}
\usepackage{xcolor}  
\usepackage{refcount}

\begin{document}

\title{Sustained photon pulse revivals from inhomogeneously broadened spin ensembles}

\author{Dmitry O. Krimer}
\email[]{dmitry.krimer@gmail.com}
\affiliation{Institute for Theoretical Physics, Vienna University of Technology (TU Wien), Wiedner Hauptstra\ss e 8-10/136, 1040 Vienna, Austria}
\author{Matthias Zens}
\affiliation{Institute for Theoretical Physics, Vienna University of Technology (TU Wien), Wiedner Hauptstra\ss e 8-10/136, 1040 Vienna, Austria}
\author{Stefan Putz}
\affiliation{Vienna Center for Quantum Science and Technology, Atominstitut, Vienna University of Technology (TU Wien), Stadionee 2, 1020 Vienna, Austria}
\affiliation{Department of Physics, Princeton University, Princeton, NJ 08544, USA}
\author{Stefan Rotter}
\affiliation{Institute for Theoretical Physics, Vienna University of Technology (TU Wien), Wiedner Hauptstra\ss e 8-10/136, 1040 Vienna, Austria}

\begin{abstract}
A very promising recent trend in applied quantum physics is to combine the advantageous features of different quantum systems into what is called ``hybrid quantum technology''. One of the key elements in this new field will have to be a quantum memory enabling to store quanta over extended periods of time. Systems that may fulfill the demands of such applications are comb-shaped spin ensembles coupled to a cavity. Due to the decoherence induced by the inhomogeneous ensemble broadening, the storage time of these quantum memories is, however, still rather limited. Here we demonstrate how to overcome this problem by burning well-placed holes into the  spectral spin density leading to spectacular performance in the multimode regime. Specifically, we show how an initial excitation of the ensemble leads to the emission of more than a hundred well-separated photon pulses with a decay rate significantly below the fundamental limit of the recently proposed ``cavity protection effect''.
\end{abstract}

\maketitle

%


%
\section{Introduction}
Various setups in cavity quantum electrodynamics (QED) have been intensively studied during the last decade with regard to their potential for enabling the storage and processing of quantum information. Particularly attractive in this context are so-called ``hybrid quantum systems'' (HQS) \cite{Xiang:2013aa,Kurizki:2015aa}, which combine the individual advantages of different quantum technologies. A major challenge for the realization of quantum information processing consists in ensuring coherent and reversible mapping of an encoded information between different elements in such systems \cite{Imamoglu1999, Rabl2006, Tordrup2008, Petrosyan2009, Verdu2009, Wesenberg2009}. A particularly attractive scenario in this context is realized based on atomic frequency combs or gradient memories in cavity or cavity-less setups \cite{Riedmatten:2008aa,Afzelius:2009aa,Usmani:2010aa,Gerasimov:2014aa,Jobez:2014aa,Moiseev:2001aa,Kraus:2006ab,Damon:2011aa,Zhang:2015aa,Zhang:2016aa} for which the information that one intends to store is emitted by the memory after the writing process in pulsed  revivals at equidistant times. One of the major bottlenecks of this technology is, however, that an inhomogeneous broadening of the atomic or spin ensemble, which plays the role of a quantum memory \cite{Duan:2001,Simon:2010}, typically leads to a relatively fast decoherence of the stored information \cite{Kubo:2010aa,Amsuss:2011aa,Sandner:2012aa}. To counteract this detrimental effect on the storage time, various techniques have been developed based, e.g., on refocusing pulses \cite{Staudt:2007aa}, gradient inversion methods \cite{Hedges:2010aa}, or preselecting the optimal spectral portion of the inhomogeneously broadened ensemble \cite{Bensky2012}. Other very recent studies propose to access long-lived dark or subradiant states in atomic or spin ensembles for efficient information storage \cite{Zhu:2014aa,Scully:2015,Guerin:2016,Krimer:2015aa,Putzprivcom}. Also new setup designs without any inhomogeneous broadening such as those based on magnon modes strongly coupled to a cavity have recently been realized \cite{Zhang:2015aa,Zhang:2016aa}. In this case, however, the gradient memory is characterized by relatively large intrinsic losses which impose limitations on the achievable time span of the revival dynamics. From these state-of-the-art experiments it is clear that new ideas and concepts will be needed to make these quantum memories viable for practical implementations, in particular in terms of the achievable storage time and the associated information retrieval efficiency.   

In this work, we propose a novel approach to obtain a sustained emission of photon pulses from spin-ensembles in spite of a significant inhomogeneous broadening of the spin transition frequencies. Our concept is not restricted to a particular experimental realization of a spin ensemble, but can instead be generally applied to different physical realizations based, for instance, on negatively charged nitrogen-vacancy (NV) defects in diamond \cite{Amsuss:2011aa,Sandner:2012aa, Kubo:2010aa,Kubo:2011aa}, or rare-earth spin ensembles \cite{Probst:2013aa,Zhong:2015aa,Jobez:2014aa}. The main requirement for our theory to be applied is that the losses exhibited by each individual constituent in the ensemble, $\gamma$, are substantially smaller as compared to the bare cavity decay rate, $\kappa$. Our key insight is that the decoherence in such hybrid quantum systems can be all but suppressed by a very non-invasive preparatory step involving the burning of a certain number of narrow holes in the comb-shaped spectral spin distribution at well-defined frequencies. Such a procedure allows us to access the corresponding collective dark state to align the system with the lowest decay scale $\gamma$, and as a result, to go beyond the fundamental limit of half the bare cavity decay rate $\kappa$ set by the recently proposed ``cavity protection effect'' \cite{Diniz:2011aa,Kurucz:2011aa,Putz:2014aa,Krimer:2014aa}. In this way we demonstrate how to sustain the pulsed emission from the ensemble during very long time intervals up to a few microseconds, achieving more than a hundred well-separated pulses. 

\vspace{-0.5cm}
\section{Results}
\begin{figure}
\centering
\includegraphics*[width=1.\linewidth]{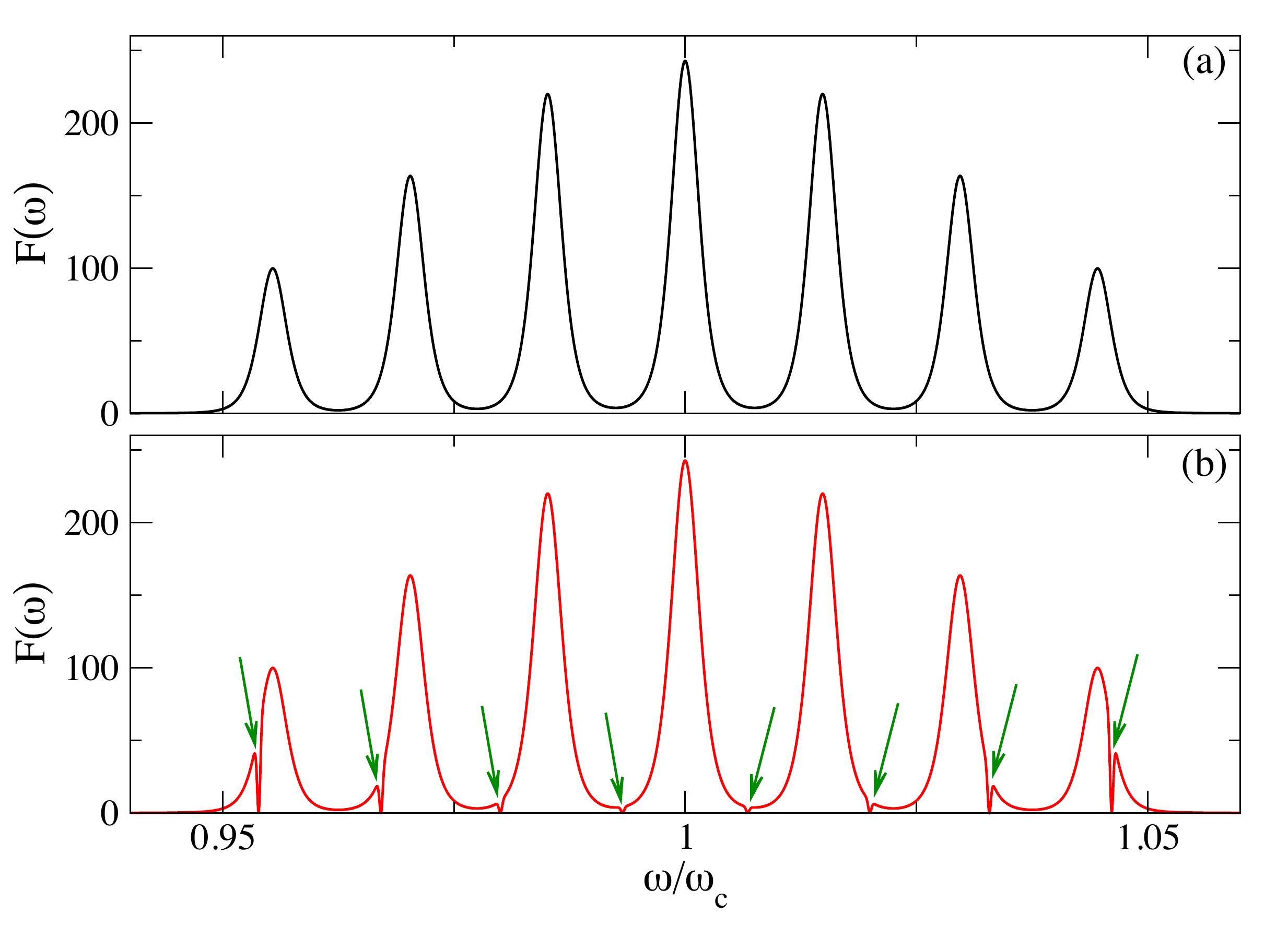}
\caption{(a) Spectral spin distribution, $F(\omega)=\sum_{\mu=1}^{M} \Omega_\mu^2/\Omega^2\cdot \rho_\mu(\omega)$, consisting of seven equally spaced $q$-Gaussians of equal width, $\gamma_q/2\pi=9.4$\,MHz. $F(\omega)$ has peaks at frequencies 
$\omega_s^{(\mu)}=\omega_c\pm n_\mu \Delta \omega$ with the spacing, $\Delta\omega/2\pi=40$\,MHz. The cavity frequency $\omega_c$ coincides with the mean frequency of the central $q$-Gaussian, $\omega_s=\omega_c=2\pi\cdot 2.6915$\,GHz. Spin ensembles have coupling strengths distributed as $\Omega_\mu^2/\Omega^2= \exp[-(\omega_c-\omega_s^{(\mu)})^2/2\sigma_G^2]$, with $\sigma_G/2\pi=150$\,MHz.  (b) Spectral function from (a) with eight spectral holes (see arrows) at the maxima of the cavity content $|A_\mu|^2$ shown in Fig.~\ref{fig_Omega_8p0}(e) for $\Omega/2\pi=26$\,MHz. All holes are of equal width,  $\Delta_h/2\pi=0.47\,$MHz, and are modelled by a Gaussian lineshape.}
\label{fig_density_new}
\end{figure}

Starting point of our analysis is an arrangement of several inhomogeneously broadened spin ensembles coupled to a single cavity mode with frequency $\omega_c$. We assume that the spin
ensembles have been prepared with mean frequencies that are equidistantly spaced at intervals of $\Delta \omega$, such that  $\omega_s^{(\mu)}=\omega_c\pm n_\mu \Delta \omega$, resulting in a comb-shaped spectral density (see Fig.~\ref{fig_density_new}a). While our approach is general we will be referring in the following to one particular experimental realization based on magnetic coupling of NV-ensembles residing in several diamonds coupled to a superconducting microwave resonator. Note that by an appropriate aligning of the diamonds with respect to an external magnetic field and by exploiting the Zeeman effect, the mean frequencies of the spin ensembles, $\omega_s^{(\mu)}$, can be efficiently tuned in a rather wide spectral interval \cite{Amsuss:2011aa,Sandner:2012aa}. To be concrete, we used in our calculations the specific parameter values from recent studies, where the non-Markovian dynamics and the cavity protection effect in a single-mode cavity strongly coupled to a single inhomogeneously broadened NV-ensemble have been studied (without holes in the spectral spin density) \cite{Krimer:2014aa,Putz:2014aa}.


%
\subsection{Theoretical model}
To account for the spin-cavity dynamics, we start from the Tavis-Cummings Hamiltonian  ($\hbar=1$) \cite{Tavis:1968aa}
\begin{eqnarray}
&&{\cal H}=\omega_ca^{\dagger}a+
\frac{1}{2}\sum_{\mu=1}^{M}\sum_{k=1}^{N_\mu}\omega_k^{(\mu)}\sigma_k^{(\mu)(z)}+
\nonumber\\
&&\text{i}\sum_{\mu=1}^{M}\sum_{k=1}^{N_\mu}\left[g_k^{(\mu)}\sigma_k^{(\mu)(-)}a^{\dagger}-g_k^{(\mu)*}\sigma_k^{(\mu)(+)}a\right]-
\nonumber\\
%
&&\text{i}\left[\eta(t) a^{\dagger}\text{e}^{-\text{i}\omega_p t}-\eta(t)^* a\text{e}^{\text{i}\omega_p t}\right]\,,
\label{Hamilt_fun}
\end{eqnarray}
where $M$ and $N_\mu$ in the summations above stand for the number of spin ensembles coupled to the single cavity mode and the number of spins in the $\mu$-th ensemble, respectively. Here $a^{\dag}$ and $a$ are standard cavity creation and annihilation operators and $\sigma_k^{(\mu)(\pm,z)}$ are the Pauli operators associated with each individual spin of frequency $\omega_k^{(\mu)}$, which obey the usual fermionic commutation relations. (The subscript $k$ enumerates an individual spin which resides in the $\mu$-th ensemble.) The interaction part of the Hamiltonian is written in the rotating-wave and dipole approximation, with $g_k^{(\mu)}$ being the coupling strength of the $k$-th spin located in the $\mu$-th ensemble. The absence of dipole-dipole interaction terms in Eq.~(\ref{Hamilt_fun}) implies that the concentration of spins in each ensemble is sufficiently low and the distance between them is large enough. The last term in Eq.~(\ref{Hamilt_fun}) describes an incoming signal with the carrier frequency $\omega_p$ and the amplitude $\eta(t)$ whose time variation is much slower as compared to $1/\omega_p$. 

Although the individual spin coupling strengths $g_k^{(\mu)}$ are very small, the effective collective coupling strength of each spin ensemble to the cavity mode, $\Omega_\mu=(\sum_{k=1}^{N_\mu} g_k^{(\mu)2})^{1/2}$, is enhanced by a factor of $\sim \sqrt{N_\mu}$. Thus, thanks to this collective coupling it becomes possible to reach the strong coupling regime by taking large ensembles (see, e.g., \cite{Amsuss:2011aa,Kubo:2011aa,Verdu2009} for NV spin ensembles). In a number of previous studies \cite{Krimer:2014aa,Putz:2014aa,Sandner:2012aa,Diniz:2011aa,Kurucz:2011aa} it was demonstrated that it is very convenient to phenomenologically introduce a continuous distribution $\rho(\omega)$ which describes the shape of the single spin spectral density. In a similar manner, we define here $M$ distributions, $\rho_\mu(\omega)=\sum_{k=1}^{N_\mu}  g_k^{(\mu)2} \delta(\omega-\omega_k^{(\mu)})/\Omega_\mu^2$, which stand for the shape of the $\mu$-th spin spectral density, each satisfying the normalization condition $\int d\omega \rho_\mu(\omega)=1$. Note that the coupling strengths $\Omega_\mu$ are not equal in general, so that the total spectral distribution acquires the following form, $F(\omega)=\sum_{\mu=1}^{M} \Omega_\mu^2/\Omega^2\cdot \rho_\mu(\omega)$, where $\Omega$ stands for the collective coupling strength of the central ensemble [see Fig.~\ref{fig_density_new}(a)]. In agreement with our previous studies \cite{Sandner:2012aa,Putz:2014aa,Krimer:2014aa}, we assume that the spectral spin density of each ensemble, $\rho_\mu(\omega)$, can be modelled by a $q$-Gaussian distribution of the following form 
\begin{equation}
\nonumber
\rho_\mu(\omega)=C\cdot\left[1-(1-q)(\omega-\omega_s^{(\mu)})^2/\Delta^2\right]^{
\dfrac{1}{1-q}},
\end{equation}
where $q$ is the dimensionless shape parameter, $1<q<3$, $\gamma_q=2\Delta\sqrt{(2^q-2)/(2q-2)}$ is the full-width at half maximum (FWHM) and $C$ is the normalization constant. 

Next, we derive the Heisenberg operator equations for the cavity and spin operators and write a set of equations for the expectation values (semiclassical approach). We consider the limit of weak driving powers and therefore the number of the excited spins is always small compared to the ensemble size. This allows us to simplify these equations by setting $\langle\sigma_k^{(\mu)(z)}\rangle\approx -1$ (Holstein-Primakoff-approximation \cite{Primakoff:1939aa}) which results in  a closed set of linear first-order ordinary differential equations (ODEs) for the cavity and spin expectation values, $A(t)=\langle a(t)\rangle$ and $B_k^{(\mu)}(t)=\langle\sigma_k^{(\mu)(-)}(t)\rangle$. Finally, by going to the continuous limit and performing rather cumbersome but straightforward calculations, we end up with a Volterra integral equation for the cavity amplitude, $A(t)=\Omega^2\int\limits_0^t d\tau {\cal K}(t-\tau) A(\tau)+{\cal D}(t)$ \cite{Krimer:2014aa}, where ${\cal D}(t)$ depends on the driving signal and initial conditions. Here the memory kernel function, ${\cal K}(t-\tau)=\int d\omega F(\omega) {\cal S}(\omega,t,\tau)$ (see Supplementary Note 1), strongly depends on the exact shape of the spectral distribution, $F(\omega)$, and is responsible for the non-Markovian feedback of the spin ensembles on the cavity, so that the cavity amplitude at time $t$ depends on all previous events $\tau<t$. (${\cal S}(\omega,t,\tau)$ depends on the time delay, $t-\tau$, frequency, $\omega$, but is independent from the spectral distribution.)

The Volterra equation turns out to be the governing equation not only for the semiclassical but also for the quantum case, when at $t=0$ the cavity is fed with a single photon and all spins in the ensembles are unexcited, $|1,\downarrow^{(i)} \rangle$. In Supplementary Note 2 we show in detail that the probability for a photon to stay inside the cavity at time $t>0$, $N(t)=\langle 1,\downarrow\!\!|a^{\dagger}(t) a(t)|1,\downarrow \rangle$, reduces to $N(t)=|A(t)|^2$ in this case, where $A(t)$ is the solution of the aforementioned Volterra equation with the initial condition $A(t=0)=1$. Note that in the context of spontaneous emission inhibition using the Zeno effect also analytical solutions of the Volterra equation have been explored \cite{Kurizki:96}.

\begin{figure}
\centering
\includegraphics*[width=1.\linewidth]{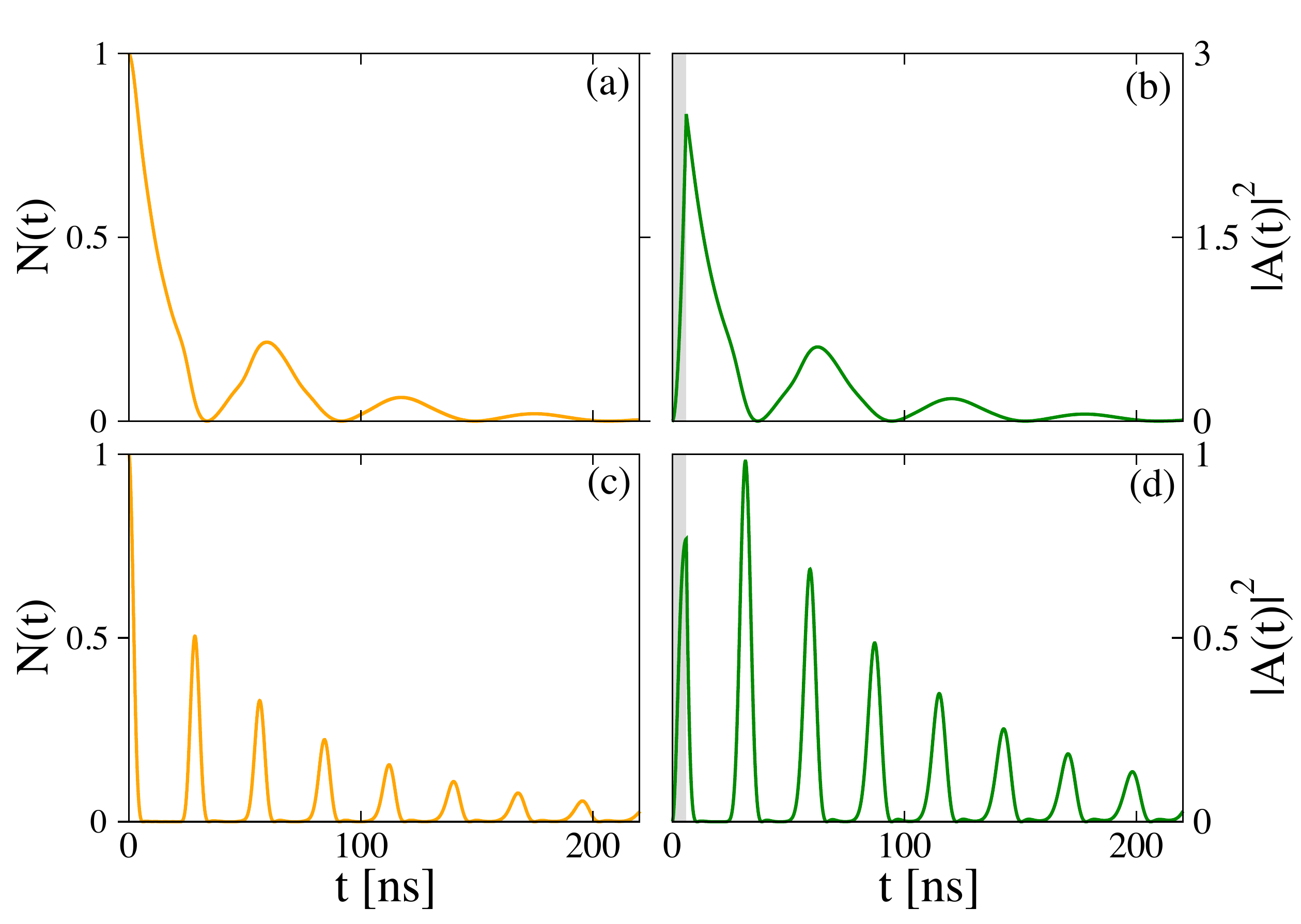}
\caption{{\it Left column}: Decay of the cavity occupation $N(t)=\,\,\,\,\langle 1,\downarrow\!\!|a^{\dagger}(t) a(t)|1,\downarrow \rangle$, when at $t=0$ the cavity is fed with a single photon of frequency $\omega_c$ and all spins are in the ground state, $|1,\downarrow \rangle$. {\it Right column}: Cavity probability amplitude $|A(t)|^2$ versus time $t$ under the action of an incident short rectangular pulse of duration $6$\,ns. The carrier frequency, $\omega_p=\omega_c=2\pi\cdot 2.6915$\,GHz. Gray (white) area indicates the time interval during which the pumping signal is on (off). (a,b) Strong coupling regime ($\Omega/2\pi=8$\,MHz) with damped Rabi oscillations. (c,d) Multimode strong coupling regime ($\Omega/2\pi=26$\,MHz) featuring pulsed revivals. The spectral function $F(\omega)$ is taken from Fig.~\ref{fig_density_new}(a) when the mean spin frequency of the central $q$-Gaussian, $\omega_s=\omega_c$ [resonant case designated by vertical cuts in Fig.~\ref{fig_Omega_8p0}(a,d)].}
\label{fig_evolution}
\end{figure}
\subsection{Multimode strong coupling dynamics} 
We first apply the Volterra equation to the spectral function $F(\omega)$ displayed in Fig.~\ref{fig_density_new}(a), for the case when the coupling strength is in the regime, where only the central spin ensemble is strongly coupled to the cavity mode (at the resonance condition $\omega_s=\omega_c$). In Fig.~\ref{fig_evolution}(a) we plot the decay of the cavity occupation $N(t)=\langle 1,\downarrow\!\!|a^{\dagger}(t) a(t)|1,\downarrow \rangle$ from the initial state, for which a single photon with frequency $\omega_c$ resides in the cavity and all spins are unexcited (the model is given in Supplementary Note 2). The resulting dynamics displays damped Rabi oscillations, which feature, however, a slightly distorted shape arising from the dispersive contribution of neighbouring spin ensembles. We observe very similar dynamics also in the semiclassical case shown in Fig.~\ref{fig_evolution}(b), when the cavity is pumped by a short rectangular microwave pulse with a carrier frequency matching the resonance condition, $\omega_p=\omega_s=\omega_c$ (see Supplementary Note 1 for the derivation of governing equations). 

In a next step, we repeat the calculations for both the quantum and the semiclassical case keeping all parameters unchanged except for the coupling strength, which we increase from $\Omega/2\pi=8$\,MHz to $\Omega/2\pi=26$\,MHz. In this limit we already entered the multimode strong coupling regime (see \cite{Krimer:2014ab}, where the reverse situation was explored, when a single emitter is coupled to many cavity modes). Correspondingly, we now observe the desired pulsed revivals of the cavity occupation $N(t)$ and the periodic emission of excitations from the spin-ensembles into the cavity amplitude $A(t)$. This type of dynamics can be attributed to a constructive rephasing of spins in the ensembles at time intervals that are  approximately equal to the inverse of the spectral distance between adjacent spin-ensembles, $2\pi/\Delta \omega$,  shown in Fig.~\ref{fig_density_new}(a). It is worth noting that we intentionally chose the duration of the initial driving pulse in Fig.~\ref{fig_evolution}(d) to be much smaller as compared to the characteristic dephasing time in our system. Such a choice ensures that the dephasing, caused by the effect of inhomogeneous broadening, only has a negligible influence up to the moment of time when the driving pulse is turned off. As a result, we obtain very regularly spaced and well-separated pulses similar to the single-photon case. As the duration of the driving pulse increases, the dephasing effect gradually sets in, and as a consequence, the dynamics becomes more and more irregular (not shown). While these results clearly show that the pulsed emission from collectively coupled and inhomogeneously broadened spin ensembles is achievable for realistic parameter values, the number of pulses that we observe in our solutions is rather limited (see Fig.~\ref{fig_evolution}). The crucial question to ask at this point is thus, whether a simple and efficient procedure can resolve this major bottleneck in the system performance.  

\subsection{Eigenvalue analysis}
 As we will show below, such a procedure can, indeed, be worked out based on a delicate modification of the spin spectral density. To arrive at this result, we need to investigate first how the eigenvalues and the corresponding eigenstates of this hybrid cavity-spin system look like. For this purpose we discretise the spectral distribution $F(\omega)$ in the frequency domain and substitute $A(t)=A\cdot \exp(-\lambda t)$ as well as $B_k^{(\mu)}(t)=B^k\cdot \exp(-\lambda t)$ into the above set of ODEs for the cavity and spin expectation values. This allows us to derive and solve numerically for each value of $\omega_s$ the non-Hermitian eigenvalue problem ${\cal L} \psi_{l}=\lambda_{l} \psi_{l}$, with $\psi_{l}=(A_l, B_l^k)^T$ being the eigenvector which represents the collective spin-cavity excitation belonging to the eigenvalue $\lambda_l$ (see Supplementary Note 4 for details). Note that Im$(\lambda_l)$ plays the role of the collective eigenfrequency and Re$(\lambda_l)>0$ is the rate at which $\psi_{l}$ decays. When solving this  eigenvalue problem we always keep the same shape for the spectral function $F(\omega)$ depicted in Fig.~\ref{fig_density_new}(a) but shift the whole structure in the frequency domain by detuning the mean spin frequency $\omega_s$ of the central ensemble with respect to the cavity $\omega_c$. [Fig.~\ref{fig_density_new}(a) corresponds to the resonant case, $\omega_s=\omega_c$.] The only other parameter that we vary is again the value of the coupling strength $\Omega$, that we tune from the limit where the cavity mode is strongly coupled solely to the 
central spin subensemble to the regime of ``multimode strong coupling''.  

\begin{figure*}
\centering
\includegraphics*[width=1.\linewidth]{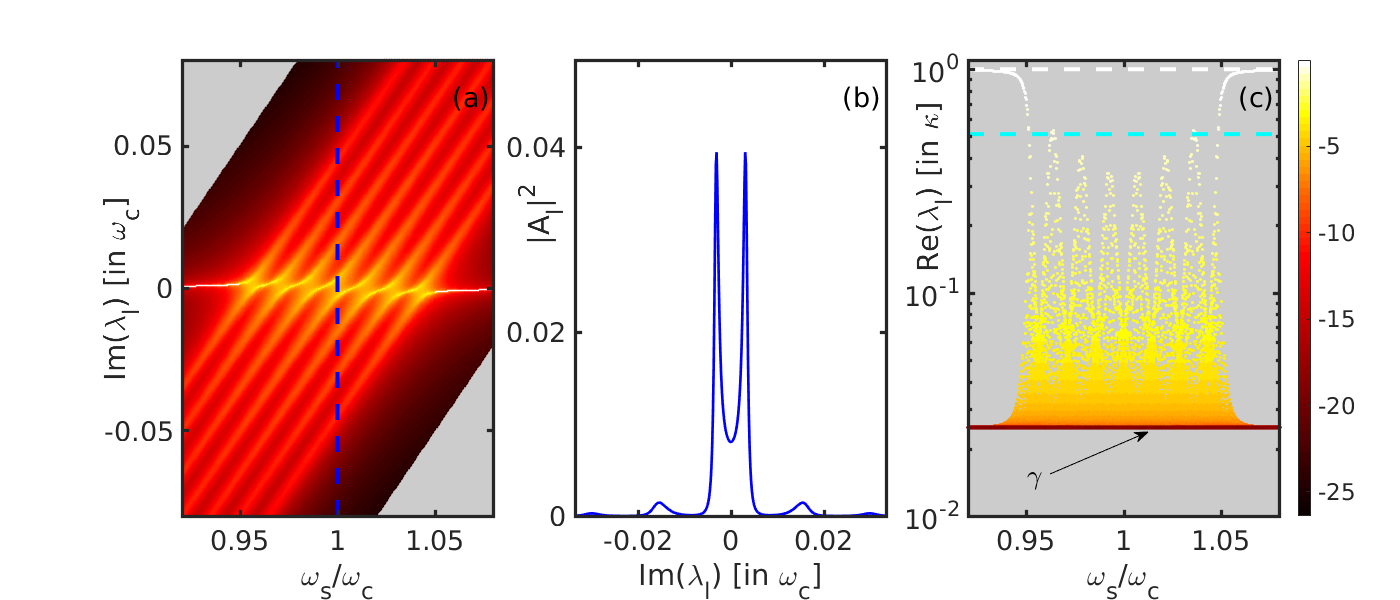}
\includegraphics*[width=1.\linewidth]{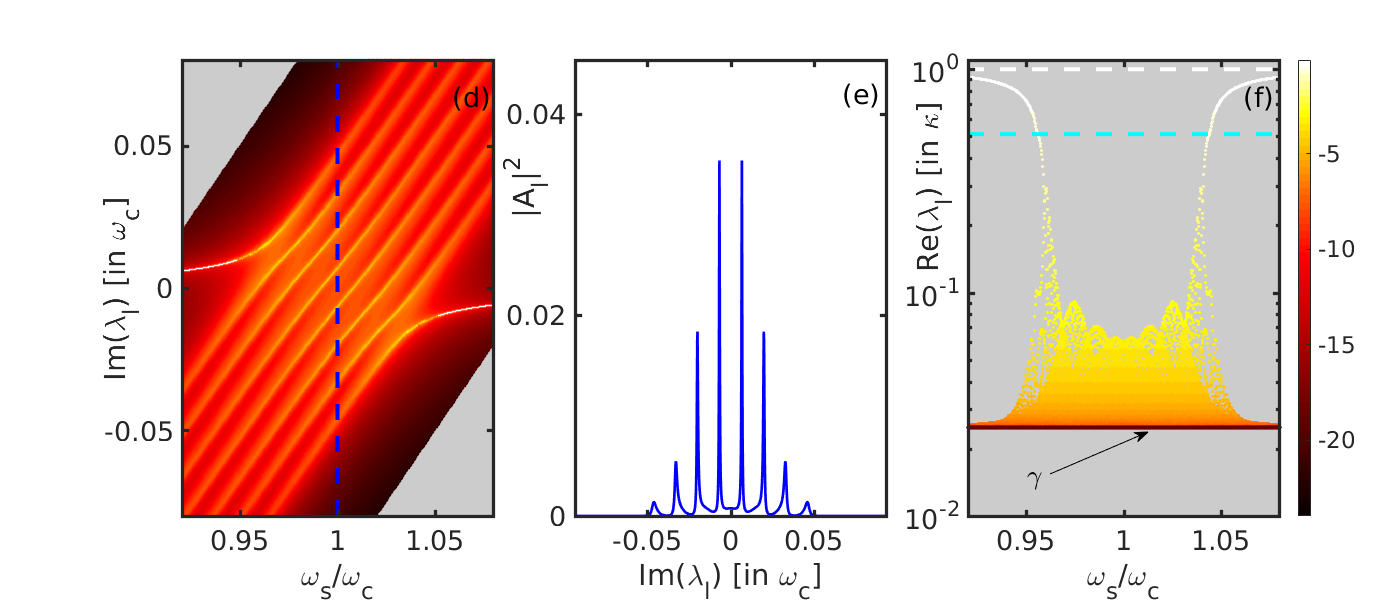}
\caption{{\it Upper row}: {\bf Single-mode strong coupling regime}. Solution of the eigenvalue problem (see the main text for details) at $\Omega/2\pi=8$\,MHz as a function of the mean spin frequency $\omega_s$ of the spectral function $F(\omega)$ shown in Fig.~\ref{fig_density_new}(a). (a) The cavity content, $|A_l|^2$, of the normalised eigenvector, $\psi_l=(A_l,B_l^k)$, versus eigenfrequencies Im$(\lambda_l)$ and $\omega_s$ is represented by the color gradient (color bar on the right in log scale): two prominent polariton modes are clearly distinguishable from a bath of dark states at fixed value of $\omega_s$. (b) the cavity content $|A_l|^2$ versus Im$(\lambda_l)$ for the resonant case, $\omega_s=\omega_c$, along the vertical cut shown in (a) (dashed blued line). (c) $|A_l|^2$ versus decay rates, Re$(\lambda_l)$, and $\omega_s$ with the same coloring as in (a). Cyan dashed line: the minimally reachable decay rate achieved due to the cavity protection effect, $\Gamma/2\approx \kappa/2$ (limit of $\gamma \ll \kappa$), with $\kappa=2\pi\cdot 0.4$\,MHz (HWHM of the cavity decay) and $\gamma=2\pi\cdot 0.01\,\text{MHz} \ll \kappa$ (HWHM of the spin decay). White dashed line: decay rate of a bare cavity mode, $\kappa$.
\newline\newline
{\it Lower row}:  {\bf Multimode strong coupling regime}. Solution of the same eigenvalue problem as above, but for an increased coupling strength $\Omega/2\pi=26$\,MHz (notation and colors are the same as in the upper row). Eight polariton modes are clearly distinguishable with an almost equidistant spacing, see (e) for the resonant case, $\omega_s=\omega_c$. In all calculations $N=1200$ spins were used.}
\label{fig_Omega_8p0}
\end{figure*}

The results of these calculations are presented in Fig.~\ref{fig_Omega_8p0}, where we plot the cavity content, $|A_l|^2$, of the normalised eigenvector $\psi_l$ as a function of $\omega_s$ and the calculated collective eigenfrequency Im$(\lambda_l)$ [(a),(d)] or decay rate Re$(\lambda_l)$ [(c),(f)].  Let us consider first the regime where the value for the coupling strength $\Omega_\mu$ of each spin ensemble separately is large enough to ensure strong coupling to the cavity. In this ``single-mode strong coupling limit'' we observe an  avoided crossing in Fig.~\ref{fig_Omega_8p0}(a) whenever  the resonance condition with the $\mu$-th ensemble is met, $\omega_s^{(\mu)}=\omega_c$. The other off-resonant spin ensembles in turn give rise to small dispersive contribution only. The most pronounced avoided crossing is observed when the cavity is at resonance with the central spin ensemble, $\omega_s=\omega_c$, where two symmetric polaritonic peaks in the structure of $|A_l|^2$ occur, see Fig.~\ref{fig_Omega_8p0}(b). It is also seen from Fig.~\ref{fig_Omega_8p0}(c) [yellow symbols] that a large fraction of eigenstates, $\psi_l$, decays with some intermediate values of the decay rate which lie within the interval $\gamma<\text{Re}\lambda_l<\kappa$. (Here $\kappa$ and $\gamma \ll \kappa$ are the dissipative cavity and spin losses,
respectively.) This can be explained by the fact that such eigenvectors represent an entangled spin-cavity state, where both the cavity and spin contents are essentially nonzero. 

With a further increase of the coupling strength, the distance between two polaritonic peaks depicted in Fig.~\ref{fig_Omega_8p0}(b), which is approximately as large as $2 \Omega$, increases and the peak line shapes become substantially sharper (not shown). Such a peak narrowing can be attributed to the so-called ``cavity protection effect'' \cite{Putz:2014aa,Kurucz:2011aa,Diniz:2011aa,Krimer:2014aa} that appears in the strong coupling regime provided that the spin density has a spectral distribution with tails that decay sufficiently fast. The latter requirement is indeed satisfied in our case because the spectral function $F(\omega)$ in Fig.~\ref{fig_density_new}(a) consists of seven $q$-Gaussian distributions.  

At even larger values of $\Omega$ the usual form of the avoided crossings eventually disappears, being replaced instead by a comb-shaped structure with parallel stripes characterised by a large cavity content, see yellow curves in Fig.~\ref{fig_Omega_8p0}(d). Such a picture is, however, valid only for moderate values of detuning of $\omega_s$ from $\omega_c$, whereas for large detuning we are in the dispersive regime [see Fig.~\ref{fig_Omega_8p0}(d,f)]. A comb-shaped structure of $|A_l|^2$ with almost equally spaced polaritonic peaks is clearly seen at resonance, $\omega_s=\omega_c$, indicating the multimode strong coupling between all spin ensembles and the cavity mode [see Fig.~\ref{fig_Omega_8p0}(e)]. It is worth noting that the peaks become substantially sharper as compared to the case of the single-mode strong coupling regime [compare Fig.~\ref{fig_Omega_8p0}(e) with Fig.~\ref{fig_Omega_8p0}(b)] due to the aforementioned ``cavity protection effect''. These narrow peaks in the frequency domain are exactly those that are responsible for the pulsed emission in the time domain as observed in Fig.~\ref{fig_evolution}(d).

The shapes of $|A_l|^2$ versus Im$(\lambda_l)$ at $\omega_s=\omega_c$ for both the multimode and the single-mode strong coupling regime reproduce exactly the corresponding shapes of the kernel function $U(\omega)$ obtained in the framework of the Laplace transform technique sketched in Supplementary Note 3 [compare Fig.~\ref{fig_Omega_8p0}(b,e) in the main text with Fig.~2 (a,b) in Supplementary Material]. The connection between these two complementary concepts provides instructive insights into the physics underlying the multimode strong coupling regime.

\begin{figure*}
\centering
\includegraphics*[width=1.\linewidth]{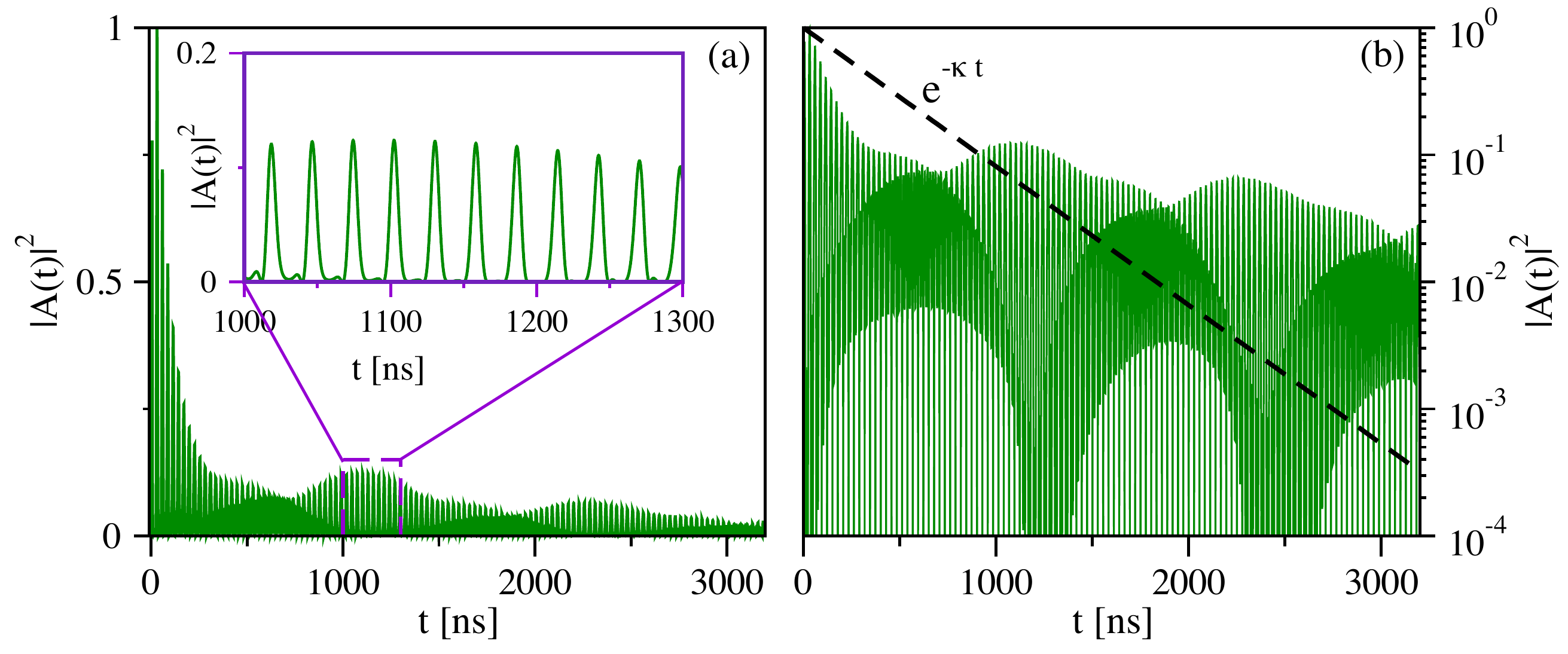}
\caption{(a) Cavity probability amplitude $|A(t)|^2$ versus time $t$ under the action of an incident short rectangular pulse of duration $6$\,ns after eight holes are burnt at $t=0$, see Fig.~\ref{fig_density_new}(b). All other parameters are the same as for the case without hole burning presented in Fig.~\ref{fig_evolution}(d). (b) Same as (a) with the ordinate plotted on a logarithmic scale. The decay process with the minimal decay rate reachable by the cavity protection effect, $e^{-\kappa t}$, with $\kappa/2\pi=0.4$\,MHz is depicted by the dashed line (limit of $\gamma \ll \kappa$). The decay rate of $|A(t)|^2$ for a bare cavity without spin ensembles coupled to it is given by $e^{-2\kappa t}$ (not shown).}
\label{fig_Gauss_0p3MHz_log_2holes_OmegaR_half}
\end{figure*}
\subsection{Suppression of decoherence}
Specifically, we want to apply these findings now to the suppression of decoherence in the multimode strong-coupling regime \cite{Krimer:2014ab,Zhang:2015aa}. For this purpose we will make use of the recent insight \cite{Krimer:2015aa}, that for single-mode strong-coupling the decoherence induced by the spin broadening can be strongly suppressed simply by burning two narrow spectral holes in the spin spectral density close to the maxima of the two polaritonic peaks as shown here in Fig.~\ref{fig_Omega_8p0}(b). The working principle of this effect is based on the creation of long-lived collective dark states \cite{Diniz:2011aa,Kurucz:2011aa,Zhu:2014aa,Putzprivcom} in the spin ensemble that only have very little cavity content and may thus even outperform the ultimate limit for the decoherence rate of the cavity protection effect given by $\Gamma=\kappa$ for $\gamma \ll \kappa$  \cite{Krimer:2015aa}. (Note that the decay rate for a bare cavity without spin ensembles coupled to it is $2\kappa$.) Mathematically, this effect can be associated with rapid variations of the nonlinear Lamb shift around the holes'  positions and with contribution of poles in the Laplace transform of the Volterra equation derived above \cite{Krimer:2015aa}. Since this theoretical concept has meanwhile also been successfully implemented in a corresponding experiment \cite{Putzprivcom}, we will try to generalize it here to the case where not just two polaritonic peaks appear (as for single-mode strong coupling), but instead many of them (as for multimode strong coupling).

The most natural extension of this hole-burning approach to the multimode regime would demand that the positions of the burned spectral holes remain close to the polaritonic peaks of which we observe altogether eight in Fig.~\ref{fig_Omega_8p0}(e), corresponding to the seven spin-subensembles shown in Fig.~\ref{fig_density_new}(a). As illustrated in Fig.~\ref{fig_density_new}(b), we therefore propose to burn eight narrow spectral holes into the spin distribution at frequencies which correspond to the maxima of the cavity content, $|A_l|^2$, shown in Fig.~\ref{fig_Omega_8p0}(e) (or, equivalently, to the maxima of the kernel function $U(\omega)$ depicted in Fig.~2(b) of the Supplementary material). The hole burning itself can be straightforwardly implemented in the experiment by exposing the cavity to very high intensity tones that feature frequency components exactly at the desired holes positions. In this way the spins at these frequency values will be shuffled into an equal population of ground and excited states, where they can no longer couple to the cavity and thus effectively form a hole in the spin distribution. This hole burning is essentially a nonlinear process, which can not be captured by the Volterra equation, but we may very well describe the system dynamics right after the holes have been burned. For this purpose we directly integrate the Volterra equation numerically in time, resulting in the time evolution for both the quantum and the semiclassical case, which looks qualitatively very similar for both cases (see Fig.~\ref{fig_Gauss_0p3MHz_log_2holes_OmegaR_half}, where the results for the semiclassical case are presented only). For these results we assume that the holes are burned at $t=0$ and that they keep their shape during the whole time interval shown in Fig.~\ref{fig_Gauss_0p3MHz_log_2holes_OmegaR_half}, a property which is well-fulfilled in recent experiments where the hole lifetime was estimated to be as large as $27\,\mu$s \cite{Putzprivcom}. Most importantly, we can see very clearly in Fig.~\ref{fig_Gauss_0p3MHz_log_2holes_OmegaR_half} that the pulsed emission from the spin ensemble persists over a drastically increased time interval as compared to the corresponding case without hole burning represented in Fig.~\ref{fig_evolution}(d). This suppression of decoherence is not only a quantitative improvement, but it breaks the barrier achievable when making maximal use of the ``cavity protection effect''. To illustrate this explicitly, we replot in Fig.~\ref{fig_Gauss_0p3MHz_log_2holes_OmegaR_half}(b) our results from Fig.~\ref{fig_Gauss_0p3MHz_log_2holes_OmegaR_half}(a) on a logarithmic scale and compare them with this minimal exponential decay $e^{-\kappa t}$ of the fully cavity-protected ensemble. We find that the probabilities $|A(t)|^2$ for the photon pulse revivals significantly exceed this barrier such that, e.g., at $t\sim 3\,\mu s$ after the driving pulse, the values for $|A(t)|^2$ are two orders of magnitude above those achievable through cavity-protection. For longer time-scales this outperformance ratio continues to grow. To check if the holes we burned in the ensemble are, indeed, located at the optimal positions, we also performed additional calculations in which we varied the hole positions by only a few percent away from the maxima of $|A_l|^2$. We find that such a shift leads to a substantial decrease in the revival amplitudes as compared to those in Fig.~\ref{fig_Gauss_0p3MHz_log_2holes_OmegaR_half} (not shown), thereby confirming our initial choice of positioning the holes right at the frequencies of the polaritonic peaks to secure the long-lived photon pulse revivals. 

\section{Conclusions and Outlook}
In conclusion, our study provides a novel approach to suppress the decoherence in quantum memories based on inhomogeneously broadened spin ensembles coupled to a cavity. Specifically, when the ensembles feature a comb-shape structure to give rise to repetitive photon pulse revivals, we show how the burning of narrow holes in this atomic frequency comb leads to a dramatic prolongation of the revival dynamics. We emphasize that the positions of the holes are generally incommensurate with the positions of the peaks in the frequency comb - a result that follows directly from our theory for the multimode strong coupling regime. Since our protocol successfully manages to overcome the decoherence both from the inhomogeneous spin broadening as well as from the cavity dissipation, we expect our protocol to be an important step towards future possible realizations of quantum memories based on spin ensembles. 

The expected challenges on the experimental side are the preparation of a comb-shaped spectral spin distribution (e.g., by detuning several sub-ensembles from each other) as well as the strong coupling to a single-mode cavity. Following our proposal, several narrow spectral holes then need to be burned into this spin ensemble (through the cavity or from the outside). After such a preparatory step, the quantum information (as stored, e.g., in a qubit \cite{Kubo:2011aa}) may be transferred through the cavity bus to the spins from where it is reemitted back into the cavity at periodic time intervals without requiring any further control or refocusing techniques.

\section*{acknowledgement}
We would like to thank A. Angerer, R. Glattauer,  B. Hartl, M. Liertzer, J. Majer, W. J. Munro and J. Schmiedmayer for helpful discussions and acknowledge support by the Austrian Science Fund (FWF) through Project No.~F49-P10 (SFB NextLite). S.P. acknowledges support by the Austrian Science Fund (FWF) in the framework of the Doctoral School "Building Solids for Function" (Project W1243).

\begin{widetext}
\newpage
{\bf Supplementary Note 1. Volterra equation for the cavity amplitude}
\vspace{0.5cm}
%

Our starting point is the Hamiltonian (1) of the main article from which we derive the equations for the cavity and spin operators, $\dot a=i [{\cal H},a]$, $\dot \sigma_k^{(\mu)(-)}=i[{\cal H},\dot \sigma_k^{(\mu)(-)}]$, respectively. Here $a$ stands for  the cavity operator and $\sigma_k^{(\mu)(-)}$ are standard Pauli operators associated with the $k$-th spin residing in the $\mu$-th ensemble. (All notations are in tact with those introduced in the main article.)  During the derivations we use the following simplifications and approximations valid for various experimental realizations: (i) $kT\ll\hbar \omega_c$ (the energy of photons of the external bath is substantially smaller than that of cavity photons); (ii) the number of microwave photons in the cavity remains small as compared to the total number of spins participating in the coupling (limit of low input powers of an incoming signal), so that the Holstein-Primakoff-approximation, $\langle \sigma_k^{(\mu)(z)} \rangle \approx -1$, always holds; (iii) the effective collective coupling strength of each spin ensemble, $\Omega_\mu^2=\sum_{k=1}^{N_\mu} g_k^{(\mu)2}$, satisfies to the inequality $\Omega_\mu \ll \omega_c$, justifying the rotating-wave approximation; (iv) the spatial size of the spin ensembles is sufficiently smaller than the wavelength of a cavity mode. Having introduced all these assumptions, we derive the following system of coupled first-order linear ordinary operator equations for the cavity and spin operators in $\omega_p=\omega_c$-rotating frame
\begin{subequations}
\begin{eqnarray}
\label{Eq_a_Volt}
&&\dot{a}(t)= -\kappa \cdot a(t)+ 
\sum_{\mu=1}^{M}\sum_{k=1}^{N_\mu} g_k^{(\mu)}\sigma_k^{(\mu)(-)}(t)\!-\!\eta(t), \\
\label{Eq_bk_Volt}
&&\dot{\sigma}_k^{(\mu)(-)}(t)\! =\! -\left[\gamma\!+\!i(\omega_k^{(\mu)}-\omega_c) \right]\sigma_k^{(\mu)(-)}(t) - g_k^{(\mu)} a(t),
\end{eqnarray}
\end{subequations}
where $\kappa$ and $\gamma$ are the total dissipative cavity and individual spin losses. By formally integrating the equations (\ref{Eq_bk_Volt}) for the spin operators and inserting them into Eq.~(\ref{Eq_a_Volt}) for the cavity operator, we get
\begin{eqnarray}
\nonumber
\dot a(t)=-\kappa \cdot a(t)+
\!\sum_{\mu=1}^{M}\sum_{k=1}^{N_\mu}   g_k^{(\mu)} \sigma_k^{(\mu)(-)}(0) e^{-i\left(\omega_k^{(\mu)}-\omega_c-i\gamma\right)t}-\Omega^2 \int_0^{\infty} \!\!d\omega F(\omega) \int\limits_{0}^t d\tau e^{-i(\omega-\omega_c-i\gamma)(t-\tau)}a(\tau)-\eta(t),
\label{Eq_a_with_Bk0}
\end{eqnarray}
where $\sigma_k^{(\mu)(-)}(0)$ is the initial spin operator and $F(\omega)$ stands for the total spectral function which is defined as a sum over the spectral densities of each spin ensemble, $F(\omega)=\sum_{\mu=1}^{M} \Omega_\mu^2/\Omega^2\cdot \rho_\mu(\omega)$. Here $\rho_\mu(\omega)=\sum_{k=1}^{N_\mu}  g_k^{(\mu)2} \delta(\omega-\omega_k^{(i)})/\Omega_i^2$ describes the spin spectral density of the $\mu$-th ensemble, $\Omega_\mu=(\sum_{k=1}^{N_\mu} g_k^{(\mu)2})^{1/2}$ is its effective collective coupling strength to the cavity mode and $\Omega$ stands for the coupling strength of the central ensemble. 

We then treat the problem semiclassically by introducing the cavity and spin expectation values, $A(t)=\langle a(t)\rangle$ and $B_k^{(\mu)}(t)=\langle\sigma_k^{(\mu)(-)}(t)\rangle$. For the sake of simplicity we consider the case when all spins are initially in the ground state, $B_k^{(\mu)}(0)=0$, so that Eq.~(\ref{Eq_a_with_Bk0}) reduces to the closed Volterra integro-differential equation for the cavity amplitude
\begin{eqnarray}
\dot A(t)=-\kappa \cdot A(t)-\Omega^2 \int_0^{\infty} \!\!d\omega F(\omega) \int\limits_{0}^t d\tau e^{-i(\omega-\omega_c-i\gamma)(t-\tau)}A(\tau)-\eta(t).
\label{Eq_a_with_Bk0_II}
\end{eqnarray}
Next we formally integrate Eq.~(\ref{Eq_a_with_Bk0_II}) in time and simplify the resulting double integral on the right-hand side by means of the partial integration method. Assuming that the cavity is initially empty, $A(0)=0$, we finally derive the following Volterra integral equation for the cavity amplitude
\begin{eqnarray}
\label{Volt_eq}
A(t)=\int\limits_0^t d\tau {\cal K}(t-\tau) A(\tau)+{\cal D}(t),
\end{eqnarray}
where ${\cal K}(t-\tau)$ is the kernel function
\begin{eqnarray}
\label{Volt_eq_K}
{\cal K}(t-\tau)=\Omega^2 \bigintsss\!\!\! d\omega\,
\dfrac{F(\omega) \left[e^{-i (\omega-\omega_c-i (\gamma-\kappa))(t-\tau)}-1\right]
}{i (\omega-\omega_c-i (\gamma-\kappa))}\cdot e^{-\kappa(t-\tau)},
\end{eqnarray}
and the function ${\cal D}(t)$ is given by
\begin{eqnarray}
\label{Volt_eq_F}
{\cal D}(t)=\int\limits_0^t d\tau\, \eta(\tau)\cdot e^{-\kappa(t-\tau)}.
\end{eqnarray}
We solve then Eq.~(\ref{Volt_eq}) numerically using the methods described in details in our recent publications \cite{KPMS14,Nature2014}.

\vspace{0.5cm}
{\bf Supplementary Note 2. Single-photon dynamics}
\vspace{0.5cm}

Here we prove that the probability for a single photon, which is populating the cavity at time $t=0$, to stay inside the cavity at $t>0$, reduces to $N(t)=|A(t)|^2$, where $A(t)$ is the solution of the Volterra equation (\ref{Eq_a_with_Bk0_II}) with the initial condition $A(0)=1$ and $\eta(t)=0$. By definition, this probability is nothing more than the expectation value of the number operator ${\cal N}=a^{\dagger}(t) a(t)$, i.e. $N(t)=\langle 1,\downarrow\!\!|a^{\dagger}(t) a(t)|1,\downarrow \rangle$. Taking into account that we deal with a single excitation in the system, we make use of the following closure relation,
\begin{eqnarray}
\mathds{1}=|0,\downarrow \rangle \langle 0, \downarrow\!\!|+\sum_l |0,\uparrow_l \rangle \langle 0, \uparrow_l\!\!|+|1,\downarrow \rangle \langle 1, \downarrow\!\!|+\sum_l |1,\uparrow_l \rangle \langle 1, \uparrow_l\!\!|,
\end{eqnarray}
where for the sake of notational simplicity the index $l$ enumerates all spins independently of the spin ensemble to which they belong to. We derive the following expression for $N(t)$
\begin{eqnarray}
\label{Eq_Nt}
&&N(t)\equiv \langle 1,\downarrow\!\!|a^{\dagger}(t) \mathds{1} a(t)|1,\downarrow \rangle=
\\\nonumber
\\\nonumber
&&|\langle 0, \downarrow\!\!| a(t)|1,\downarrow \rangle|^2+|\langle 1, \downarrow\!\!| a(t)|1,\downarrow \rangle|^2+\sum_l |\langle 0, \uparrow_l\!\!| a(t)|1,\downarrow \rangle|^2+\sum_l |\langle 1, \uparrow_l\!\!| a(t)|1,\downarrow \rangle|^2.\,\,\,\,\,
\end{eqnarray}
We then let the operator equations (\ref{Eq_a_Volt}, \ref{Eq_bk_Volt}) from Supplementary Note 1 act on the bra- and ket-vectors which show up in Eq.~(\ref{Eq_Nt}) and derive four independent sets of coupled ODEs for the corresponding expectation values $\langle a(t)\rangle$ and $\langle \sigma_j^-(t)\rangle$. Remarkably, these sets of equations look formally the same being independent of the specific bra- or ket-vector appearing on the left or right side in these operator equations i.e. they evolve in the same fashion as the corresponding operators themselves. The only difference between the resulting solutions for the expectation values appearing in Eq.~(\ref{Eq_Nt}) stems from the initial conditions which are nonzero only for the first term in the r.h.s. of Eq.~(\ref{Eq_Nt}), namely $A(0)=\langle 0, \downarrow\!\!| a(0)|1,\downarrow \rangle=1$. For all other terms the resulting expectation values are zero at $t=0$, and as a consequence, they remain zero at $t>0$ as well. Therefore, the probability for a photon to reside in the cavity at $t>0$ reduces to $N(t)=|\langle 0, \downarrow\!\!| a(t)|1,\downarrow \rangle|^2=|A(t)|^2$, where $A(t)$ is exactly given as the solution of the Volterra equation (\ref{Eq_a_with_Bk0_II}) in Supplementary Note 1 with the initial conditions $A(0)=1$ and $B_l(0)=0$. 

\vspace{0.5cm}
{\bf Supplementary Note 3. Laplace transform of the Volterra equation}
\vspace{0.5cm}
\begin{suppfigure}
\centering
\includegraphics*[width=.25\linewidth]{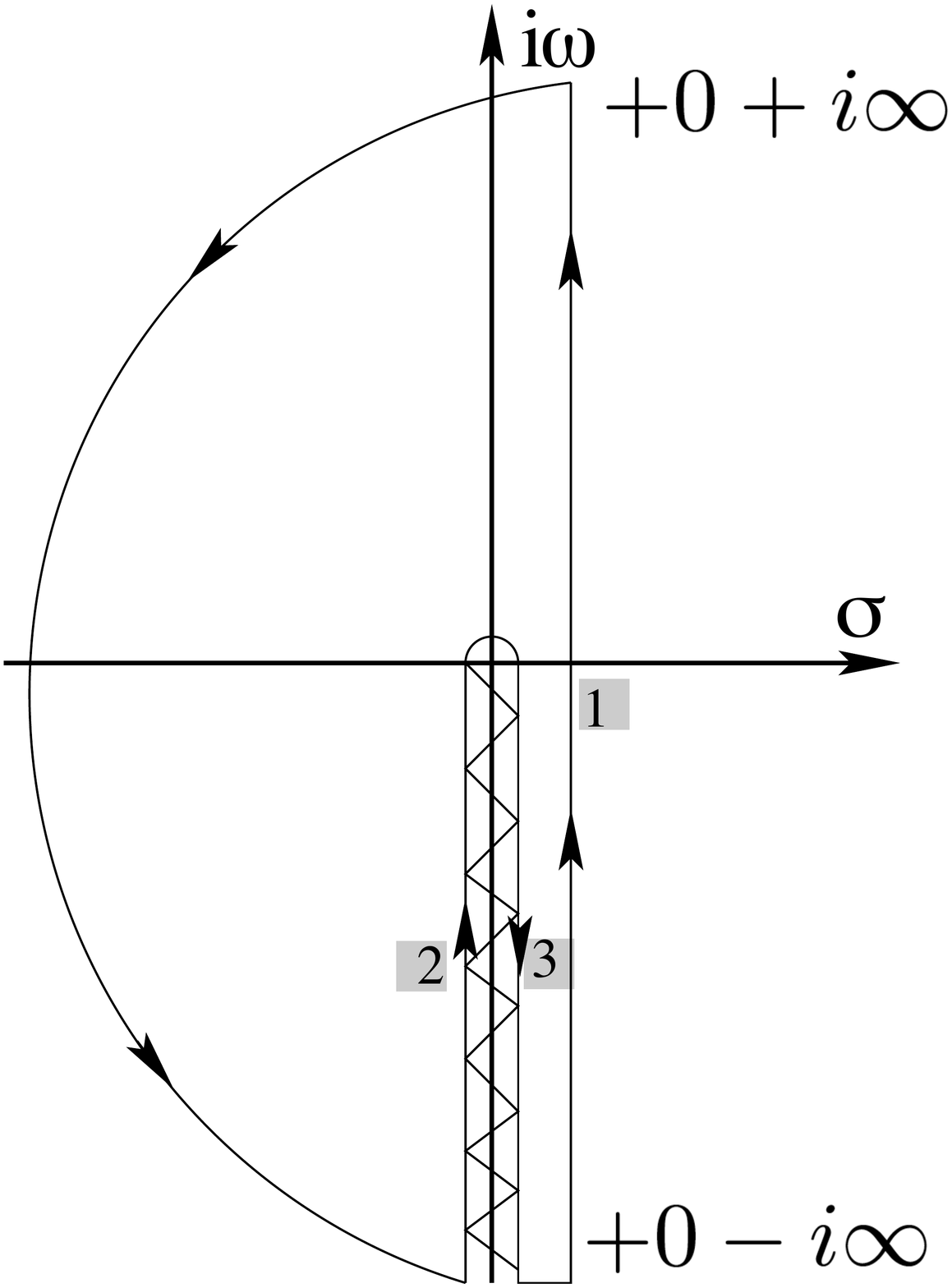}
\caption{Contour completion in the complex plane $s=\sigma+i \omega$ for the calculation of the inverse Laplace transform. Those contours which give nonzero contribution are designated by numbers. The zig-zag line corresponds to the branch cut along the negative part of the imaginary axis.}
\label{fig1}
\end{suppfigure}

Here we sketch the derivation of the Laplace transformation of the Volterra equation (\ref{Eq_a_with_Bk0_II}) from Supplementary Note 1 assuming that all spins are initially in the ground state and the cavity mode $a$ contains initially a single photon, $A(0)=1$ (the case considered in Supplementary Note 2). For that purpose we multiply Eq.~(\ref{Eq_a_with_Bk0_II}) by $e^{-st}$ ($s=\sigma+i\omega$ is the complex variable), integrate both sides of the equation with respect to time and finally obtain the following expression for the Laplace transform:
\begin{equation}
\tilde A(s)=\dfrac{1}{s+\kappa-\gamma+\Omega^2 \int_0^{\infty}
\dfrac{d\omega F(\omega)}{s+i(\omega-\omega_c)}}.
\end{equation}
By performing the inverse Laplace transformation, $A(t)=(2\pi i)^{-1}\int_{\sigma-i \infty}^{\sigma+i \infty}ds\, e^{st} \tilde A(s)$ (see e.g. \cite{Riley} for more details), we get the formal solution for the cavity amplitude $A(t)$ which is as follows
\begin{eqnarray}
\label{Inverse_LT}
A(t)\!=\!\dfrac{e^{i(\omega_c-i\gamma) t}}{2\pi i}
\int_{\sigma-i \infty}^{\sigma+i \infty} \dfrac{e^{st}ds}{s+\kappa-\gamma+i\omega_c+\Omega^2 \int_0^{\infty}
\dfrac{d\omega F(\omega)}{s+i\omega}},\,\,\,\,
\end{eqnarray}
where $\sigma>0$ is chosen such that the real parts of all singularities of $\tilde A(s)$ are smaller than $\sigma$. It turned out that the integral in the denominator of Eq.~(\ref{Inverse_LT}) has a jump when passing across the negative part of the imaginary axis leading to the branch cut in the complex plane of $s$ (see Fig.~\ref{fig1}). 
By setting the denominator of the integrand in Eq.~(\ref{Inverse_LT}) to zero, one can derive the equations for simple poles, $s_j=\sigma_j+i\omega_j$, which, however, do not appear for the spectral function shown in Fig.~1(a) of the main paper and will not be discussed here (see \cite{KPMS14} for more details about poles' contribution).

\begin{suppfigure}
\centering
\includegraphics*[width=.85\linewidth]{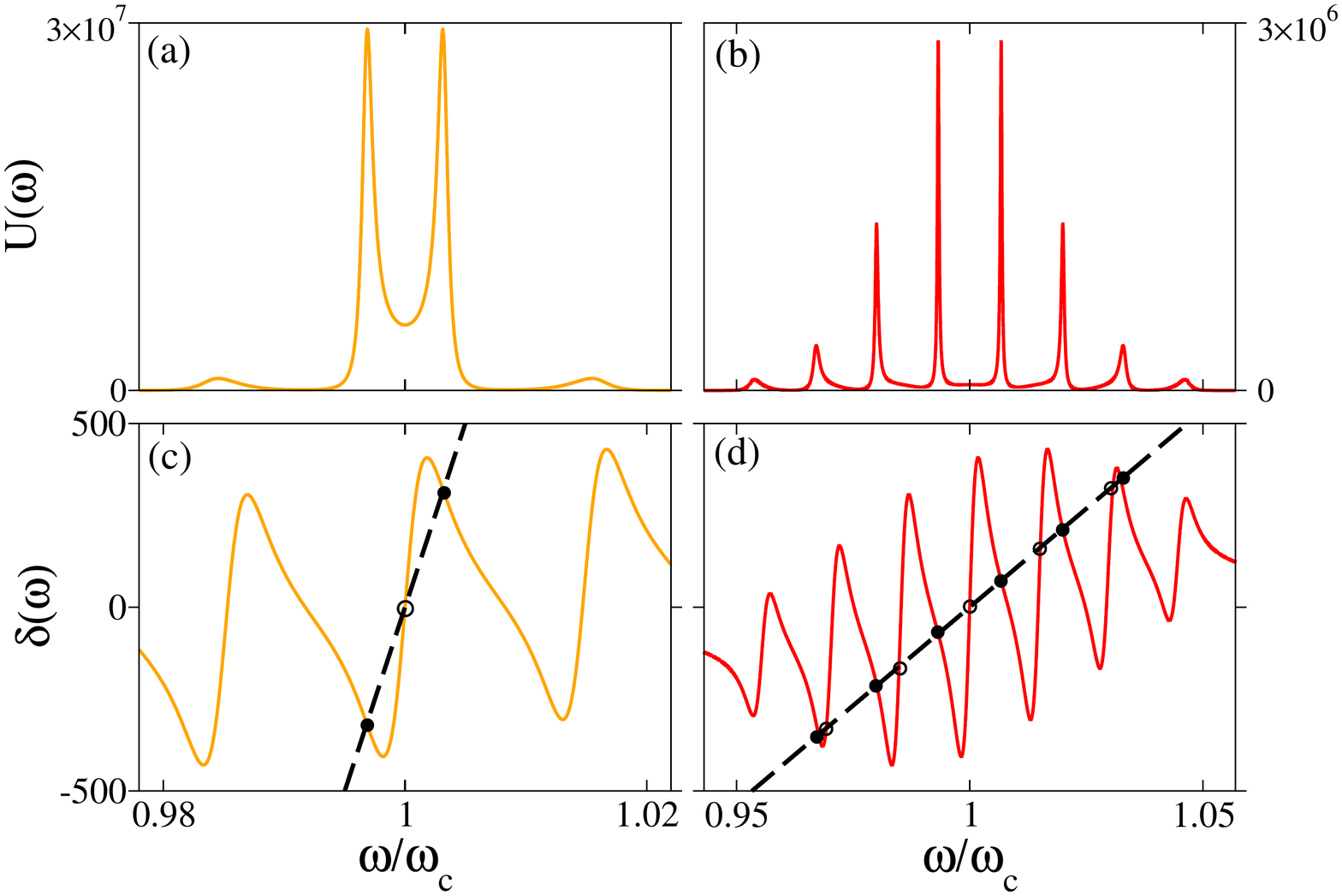}
\caption{Route from strong coupling to multimode strong coupling regime for two different coupling strengths, $\Omega/2\pi=8$\,MHz (left column) and $\Omega/2\pi=26$\,MHz (right column). {\it Upper row}: Kernel function $U(\omega)$. {\it Lower row}: Nonlinear Lamb shift $\delta(\omega)$ for the same $\omega$-interval as above (note the different zooms for the two columns). {\it Left column}: Strong coupling regime with a well-resolved Rabi splitting in $U(\omega)$ (regime of damped Rabi oscillations).  {\it Right column}: Multimode strong coupling regime with a multi-peak structure in $U(\omega)$ when all seven spin ensembles are effectively coupled to the cavity (regime of revivals). Filled circles label resonance values $\omega_r$ of the kernel $U(\omega)$ occurring at the intersections between the Lamb shift $\delta(\omega)$ and the dashed line $(\omega-\omega_c)/\Omega^2$. At empty circles such intersections are non-resonant and do not lead to a corresponding peak in $U(\omega)$. The cavity frequency $\omega_c$ coincides with the mean frequency of the central $q$-Gaussian, $\omega_s=\omega_c$, shown in Fig.~1(a) of the main article.}
\label{fig_spectrum}
\end{suppfigure}

Next, we apply Cauchy's theorem to a closed contour to evaluate the formal integral (\ref{Inverse_LT}) taking into account that only a few paths of those shown in Fig.~S\ref{fig1} contribute. Finally, we end up with the following expression for the cavity amplitude
\begin{eqnarray}
A(t)= \Omega^2\int_{0}^{\infty} d\omega e^{-i(\omega-\omega_c-i\gamma) t} U(\omega),\,\,\,\,\,
\end{eqnarray}
where
\begin{eqnarray}
\label{Eq_Phis_34}
U(\omega)=\lim_{\sigma\rightarrow 0^{+}}\left\{\dfrac{F(\omega)}
{\left(\omega\!-\!\omega_c\!-\!\Omega^2 \delta(\omega)+\!i(\kappa-\gamma)\right)^2\!+\!(\pi\Omega^2F(\omega)\!+\!\sigma)^2}\right\}.
\end{eqnarray}
is the kernel function and 
\begin{eqnarray}
\delta(\omega)=\mathcal{P}\int_0^{\infty}\dfrac{d\tilde\omega F(\tilde\omega)}{\omega\!-\!
\tilde\omega}\!
\label{Eq_Lamb_shift}
\end{eqnarray}
has the meaning of the nonlinear Lamb shift of the cavity frequency $\omega_c$, which depends on the total spectral distribution, $F(\omega)$. 

Obviously, the relevant frequency components contributing to the dynamics of $A(t)$ are those which are resonant in the kernel function $U(\omega)$. As it can be deduced from the structure of $U(\omega)$ given by Eq.~(\ref{Eq_Phis_34}), a necessary condition for such resonances to show up strongly depends on the structure of the Lamb shift and the value of the coupling strength. Namely, it is given by the following approximate formula, $(\omega_r-\omega_c)/{\Omega^2} \approx \delta( \omega_r)$. At small values for the coupling strength $\Omega$ the straight line $(\omega_r-\omega_c)/{\Omega^2}$ becomes very steep and thus leads just to a single intersection with $\delta( \omega_r)$. As a result, a single resonance occurs at $\omega_r \approx \omega_c$, so that only the central spin ensemble contributes to the coupling with the cavity, whereas the others yield a negligible contribution. In this case the kernel function $U(\omega)$ can be well approximated by a Lorentzian centered around the slightly shifted cavity frequency $\omega_c+\Omega^2 \delta(\omega_c)$.  We will thus deal with the exponential decay of the cavity amplitude $A(t)$ in the time domain with a decay rate depending on $\Omega$. Actually this regime is very similar to the Purcell enhancement of the spontaneous emission rate of a single emitter inside a cavity \cite{Purcell}. As the coupling strength reaches a certain critical value, the straight line intersects the nonlinear Lamb shift at three points from which only two give rise to the resonances in $U(\omega)$. As a consequence, the kernel function $U(\omega)$ consists of two well-separated polaritonic peaks, which is the hallmark of the strong coupling regime of cavity QED (see the left column in Fig.~S\ref{fig_spectrum}). Note that these two resonances still reside in the vicinity of the cavity frequency and the contribution of all but the central ensemble is rather small. 

The situation changes qualitatively at higher values of the coupling strength $\Omega$, when the straight line also intersects the other distant resonances of the Lamb shift, as is seen from the right column in Fig.~S\ref{fig_spectrum}. As a result, the kernel function $U(\omega)$ forms a comb-shaped structure with almost equally spaced polaritonic peaks at frequencies which are shifted with respect to the resonances of the spectral function $F(\omega)$. It is worth noting, that such a communication of the cavity mode with distant resonances of $F(\omega)$ would never take place if the Lamb shift were approximated by its value at the cavity frequency, $\delta(\omega_c)$.

\vspace{0.5cm}
{\bf Supplementary Note 4. Eigenvalue problem}
\vspace{0.5cm}

To solve the eigenvalue problem we first discretize the spectral function $F(\omega)$ (see Fig.~1 in the main article) by performing the following transformation:
\begin{eqnarray}
\label{Eq_g_l}
g_l=\left[F(\omega_l)  \cdot  (\sum_{\mu=1}^{M}\Omega_\mu^2)/\sum_m F(\omega_m)\right]^{1/2}\!\!\!\!\!.
\end{eqnarray}
Since in total we deal with a sizeable number of spins, we make our problem numerically tractable by dividing spins into many subgroups with approximately the same coupling strengths, so that $g_l$ in Eq.~(\ref{Eq_g_l}) represents a coupling strength within each subgroup rather than an individual coupling strength. Note also that once a shape of the spectral function $F(\omega)$ is defined, it is not relevant anymore to which ensemble an individual  spin belongs to. By doing so we get the following linear set of first-order ODEs with respect to the cavity and spin amplitudes from Eqs.~(\ref{Eq_a_Volt},\ref{Eq_bk_Volt}) ($\eta(t)=0$)
\begin{subequations}
\begin{eqnarray}
\label{Eq_a_Volt_simpl}
&&\dot{A}(t)= -\kappa\cdot A(t) + \sum_l g_l  B_l(t) \\
\label{Eq_bk_Volt_simpl}
&&\dot{B}_l(t) = -\left[\gamma+i(\omega_l-\omega_c) \right]B_l(t) - g_l A(t),
\end{eqnarray}
\end{subequations}
where $A(t)\equiv \langle a(t)\rangle$ and $B_l(t)\equiv\langle\sigma_l^-(t)\rangle$. After substituting $A(t)=A\cdot \exp(-\lambda t)$ and $B_l(t)=B_l\cdot \exp(-\lambda t)$ into Eqs.~(\ref{Eq_a_Volt_simpl}, \ref{Eq_bk_Volt_simpl}), we derive the complex eigenvalue problem for $\lambda$, which can be represented as, ${\cal L} \psi=\lambda \psi$, where
 \begin{equation}
      {\cal L} =
      \left(\begin{array}{c c c c}
\kappa & -g_1                                            &\,\,\, -g_2  \,\,\,\,\,\,\,\,\,\,\,\,\,\,\,\,\,\,\,\,\,\,\,\,\,\,\,\,\,...    & -g_N \\
g_1      & \gamma+i(\omega_1-\omega_c) & \,\,\,\,\,\,\,\,\, 0         \,\,\,\,\,\,\,\,\,\,\,\,\,\,\,\,\,\,\,\,\,\,\,\,\,\,\,\,\,\, ...         &  0 \\
g_2      & 0                 & \gamma+i(\omega_2-\omega_c)  \,\,\,\,\,\, ...  & 0 \\
    ...      &   ...      &     \,\,\,\,\,\, \,\,  ... \,\,\,\,\,\,\,\,\,\,\,\,\,\,\,\,\,\,\,\,\,\,\,\,\,\,\,\,\,... &... \\
 g_N     & 0  & \,\,\, \,\,\,   \,\,\,0      \,\,\,\,\,\,\,\,\,\,\,\,\,\,\,\,\,\,\,\,\,\,\,\,\,\,\,\,\,\, ...  &  \gamma+i(\omega_N-\omega_c)\\  
  \end{array}\right) \mbox{,} \nonumber
 \end{equation}
and $\psi=(A\,\,\,\,\, B_1 \,\,\,\,\,  B_2 \,\,...\,\, B_N)^T$.

%


\end{widetext}
%

\end{document}